\documentclass{appolb}
\usepackage{epsfig}
\begin{document}
\title{Strangeness in the cores of neutron stars
\thanks{Strangeness in Quark Matter SQM2011, Cracow, September 18-24, 2011}
}
\author{
R.~{\L}astowiecki$^1$\footnote{\emph{e-mail address}: 
lastowiecki@ift.uni.wroc.pl}, 
D.~Blaschke$^{1,2}$\footnote{\emph{e-mail address}: blaschke@ift.uni.wroc.pl}, 
H.~Grigorian$^3$\footnote{\emph{e-mail address}: hovik.grigorian@gmail.com}, 
S.~Typel$^4$\footnote{\emph{e-mail address}: s.typel@gsi.de}
\address{
$^1$Instytut Fizyki Teoretycznej, Uniwersytet Wroc{\l}awski, Wroc{\l}aw,
Poland\\
$^2$Bogoliubov Laboratory for Theoretical Physics, JINR Dubna, Dubna, Russia\\
$^3$Department of Physics, Yerevan State University, Yerevan, Armenia\\
$^4$Nuclear Astrophysics Virtual Institute, 
GSI Helmholtzzentrum f\"ur Schwerionenforschung, Darmstadt, Germany}
}
\maketitle
\begin{abstract}
The measurement of the mass $1.97\pm 0.04~M_\odot$ for PSR J1614-2230 provides 
a new constraint on the equation of state and composition of matter at high 
densities.
In this contribution we investigate the possibility that the dense cores of 
neutron stars could contain strange quarks either in a confined state  
(hyperonic matter) or in a deconfined one (strange quark matter) while 
fulfilling a set of constraints including the new maximum mass constraint.
We account for the possible appearance of hyperons within an extended version  
of the density-dependent relativistic mean-field model, including the $\phi$ 
meson interaction channel. Deconfined quark matter is described by the color
superconducting three-flavor NJL model.
\end{abstract}
  
\section{Introduction}
The measurement of the mass $1.97\pm 0.04~M_\odot$ for PSR J1614-2230 by 
Demorest et al. \cite{Demorest:2010bx} has renewed the interest in the question
of the internal structure of a neutron star (NS).
This concerns in particular the question of its composition and the possibility
of exotic forms of matter in the cores of NSs which  has been tested 
extensively against this new measurement 
\cite{Ozel:2010bz,Lattimer:2010uk,Weissenborn:2011qu,Bonanno:2011ch}, 
thus reviving an old controversy \cite{Ozel:2006bv,Alford:2006vz}
in the course of which it had been shown that hybrid stars with 
quark matter cores could 
%be sufficiently massive to 
not be ruled out by the
observation of a $2~M_\odot$ compact star \cite{Klahn:2006iw}.

This was demonstrated not only within extended bag models 
\cite{Ozel:2010bz,Weissenborn:2011qu}
but also within the field theoretical Nambu--Jona-Lasinio (NJL) model 
description of quark matter \cite{Klahn:2006iw,Klahn:2011fb}.
In the latter approach it was found that the hybrid star becomes unstable as
soon as strange quark matter (in its color superconducting CFL phase) appears 
in the very core of the hybrid star. 
If one combines the NJL-model description of color superconducting quark matter
with an additional (possibly density dependent) bag pressure, then it is 
possible to achieve stability of hybrid stars even with strange quark
matter interior \cite{Pagliara:2007ph}, eventually fulfilling the $2~M_\odot$
constraint from PSR J1614-2230 
\cite{Ippolito:2007hn,Bonanno:2011ch,Blaschke:2010vj}.
There appears a new question which we would discuss here: 
Given a high-mass hybrid star contains quark matter, %under which conditions 
could this be strange quark matter?
 
A separate issue is the possible appearance of hyperons at supersaturation 
densities, before a phase transition to quark matter occurs.
It was found that the appearance of hyperons softens the equation of state 
(EoS) to the extent that sufficiently massive hybrid stars can not be 
described. 
In the Brueckner-Bethe-Goldstone (BBG) theory even the typical NS mass of 
$\sim 1.4~M_\odot$, well measured for binary radiopulsars, could not be 
reached \cite{Baldo:1999rq}.
An early deconfinement transition to bag model quark matter with a 
density-dependent bag constant was suggested as a possible solution
\cite{Baldo:2003vx}.
Within the density-dependent relativistic mean field (RMF) theory of neutron 
star matter \cite{Hofmann:2000mc} higher maximum masses than in BBG theory 
could be reached and the inclusion of repulsive $\phi$ meson interations 
provided a stiff enough EoS to support a  $2~M_{\odot}$ star with hyperonic 
interior \cite{Weissenborn:2011ut,Bonanno:2011ch}.

In the present contribution we discuss hybrid stars with both, hyperonic  
and quark matter interior in the light of the new maximum mass constraint.

\section{Model setup and results}

The hadronic equation of state is described by a RMF 
%relativistic mean-field 
model with density dependent nucleon-meson couplings
using the parametrization DD2 \cite{Typel:2009sy} that was fitted to
properties of finite nuclei. 
The model was extended to include all hyperons of the baryon octet in a 
similar spirit as in Ref.~\cite{Hofmann:2000mc}.
The couplings of the hyperons to the $\omega$ and $\rho$ mesons are obtained 
from a SU(3) rescaling of the nucleon meson couplings with an additional 
overall factor $R=0.83$ that is close to the one given in 
\cite{Hofmann:2000mc}.
The coupling of the $\sigma$ meson to the hyperons was determined
such that the hyperon potential in symmetric nuclear matter at
saturation assumes the values $U_{\Lambda}=U_{\Sigma}=-30$~MeV and
$U_{\Xi}=-21$~MeV, respectively. 
A repulsive interaction between the hyperons was
modeled by the exchange of the $\phi(1020)$ vector meson with a coupling
constant given by the $\omega$-nucleon coupling at saturation
density with a SU(3) scaling and the same reduction factor $R$ as for
the nucleon-meson couplings.

Quark matter is described by the color superconducting three-flavor NJL model 
with scalar, diquark and vector interaction \cite{Blaschke:2005uj}.
Parameters of the model are chosen from Ref.~\cite{Grigorian:2006qe} 
\footnote{This parametrization scheme has been implemented in an online tool 
developed by F.~Sandin which also corrects for a mistake in the kaon mass 
formula employed in ~\cite{Grigorian:2006qe}, see 
http://3fcs.pendicular.net/psolver} for the NJL model case with a constituent 
quark mass of $M(p=0)=367.5$ MeV.

We add a correction to the quark pressure to account for the melting of 
the gluon condensate due to the influence of dynamical quarks which appear in
the system for $\mu>\mu_c$, i.e. above the critical chemical potential 
$\mu_c= 1047$ MeV for chiral symmetry restoration.
The pressure of the quark matter subsystem is then 
$p_{\rm quark}(\mu)=p_{\rm NJL}(\mu) -  B(\mu)$,
where the additional contribution is understood as a chemical 
potential dependent bag pressure
\begin{equation}
 B(\mu) = B_0\left[\exp\left(-\frac{\mu-\mu_c}{\delta\mu}\right)-1\right]~~,
~~\mu>\mu_c~,
\end{equation}
and $B(\mu)=0$ elsewhere. 
The resulting correction to the density is
\begin{equation}
\Delta n(\mu)=- \frac{\partial B(\mu)}{\partial\mu} 
%= \frac{B_0}{\sigma}\exp\left(-\frac{\mu-\mu_c}{\sigma}\right).
=\frac{B_0+B(\mu)}{\delta\mu}~,
\end{equation}
which entails a contribution to the energy density 
\begin{equation}
\Delta \varepsilon(\mu)
=\varepsilon_{\rm quark}(\mu)- \varepsilon_{\rm NJL}(\mu)
= B(\mu)+\mu \Delta n(\mu)~.
\end{equation}
We report results for the parameters
$ B_0 = 40 $ MeV/fm$^3$, $\delta \mu = 100 $ MeV.

\begin{figure}[thb]
 \includegraphics[scale=0.45,angle=-90]{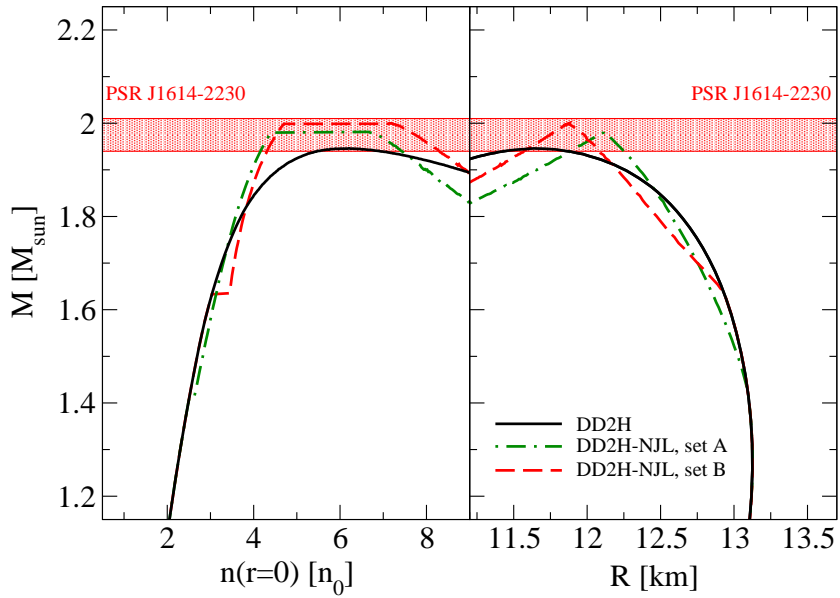}
 \caption{Mass versus central density (left panel) and versus radius (right
panel) for pure DD2 with hyperons (solid lines) and for two hybrid EoS with 
$\eta_V=0.4$: set A with $\eta_D=1.0$ (dash-dotted lines) and set B with 
$\eta_D=0.9$ (dashed lines).}
\label{FIG:MR_HYP}
\end{figure}

The hadronic-to-quark-matter phase transition is described by a Maxwell 
construction whereby the neutron star constraints are fulfilled locally, i.e., 
separately for the EoS of the two phases.
The resulting hybrid EoS is then used for solving the
Tolman-Oppenheimer-Volkoff equations. 
In Fig.~\ref{FIG:MR_HYP} we present the sequences of to hybrid star 
configurations obtained  for two sets of NJL model parameters and a purely 
hadronic EoS for comparison.
We observe that for both hybrid star sequences the maximum mass is higher 
than that for the purely hadronic case.
This result needs an explanation. 
After the first crossing of $p_{\rm hadron}(\mu)$ and $p_{\rm quark}(\mu)$,
which defines the deconfinement phase transition, there is a second crossing 
of both pressure curves due to the softness of hyperonic matter which is 
reflected in a steeper rise of  $p_{\rm hadron}(\mu)$ than of 
$p_{\rm quark}(\mu)$ at high densities. 
We disregard the second crossing by the argument that once hadrons are 
dissolved by compressing hadronic matter and thus deconfinement occurred, 
a further compression cannnot restore the hadronic phase. 
The hyperonic degrees of freedom become irrelevant and their EoS can be 
ignored beyond the first transition point. 
For a discussion of multiple crossing of pressure curves and the so-called
masquerade problem, see  \cite{Alford:2004pf}.

\begin{figure}[htb]
 \includegraphics[scale=0.45,angle=-90]{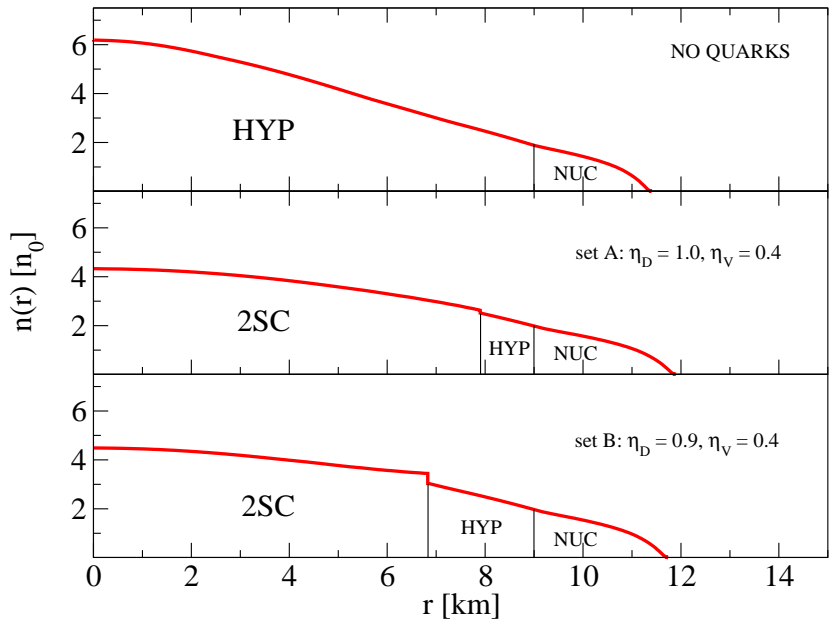}
 \caption{Structure of compact stars for a mass $1.94~M_\odot$ within the 
limits provided by PSR J1614-2230, for three cases: without quark matter 
(upper panel), with quark matter for set A %$\eta_D=1.0$ and $\eta_V=0.4$
(middle panel), and with quark matter for set B %$\eta_D=0.9$, $\eta_V=0.4$
(lower panel).}
 \label{FIG:structureHYP}
\end{figure}

In Fig.~\ref{FIG:structureHYP} we present the internal structure of the 
configurations with mass $M = 1.94~M_\odot$, which is the maximum mass 
obtained in the purely hadronic case.
We note that due to the softness of the hadronic EoS with hyperons the central 
density in the maximum mass hyperonic star exceeds $6~n_0$,
where $n_0$ is saturation density.
This configuration has an extended hyperonic inner core and a relatively 
thin ($\sim 2$ km) purely nucleonic outer core.
The two hybrid configurations with a deconfinement transition have a two-flavor
color superconducting (2SC phase) quark matter core extending up to $2/3$
of the star's radius.
Their inner core is surrounded by layers of hyperonic and purely nucleonic
matter, respectively.

We should note a certain peculiarity of the model.
Hyperons contain strangeness while the quark matter in the core is in the 
2SC phase.
Hence we have a situation where strangeness is confined to a layer of 
hyperonic matter at moderate densities, while the deconfinement phase 
transition at higher density liberates sequentially first the light flavors
and only at still higher densities also the strange quarks
\cite{Blaschke:2008br}. 
Once this occurs the matter softens and the star collapses so that the mass
at which this critical density is reached in the center is the maximum mass
for hybrid stars.

\section{Conclusion}
We conclude that one can obtain hyperon stars fulfilling the new $2~M_{\odot}$
mass constraint.
This requires the repulsive $\phi$ meson interaction channel.
Straightforward attempts to construct three-phase hybrid stars fail to meet 
the new mass constraint.
With a phenomenological $\mu$-dependent contribution to the quark pressure 
motivated by backreaction from the gluon sector a deconfinement phase 
transition could be obtained with hybrid star sequences fulfilling the new 
maximum mass constraint.
It is plain that traditional phase transition constructions introduce 
inconsistencies like a ``second crossing'' of pressure curves and should be 
replaced by a microscopic description of hadron dissociation at
high densities, to be developed.

\section{Acknowledgments}
This work has been supported in part by ``CompStar'', a Research Networking 
Programme of the European Science Foundation and by the Polish National 
Science Centre (NCN) %Ministry for Science and Higher Education 
under grant No. NN 202 2318 37.
The work of D.B. has been supported by the Russian Fund for Fundamental 
Investigations under grant No. 11-02-01538-a.
H.G. has been supported in part by the Volkswagen Foundation under grant No. 
85 182.
R.{\L}. received support from the Bogoliubov-Infeld programme for
visits at the JINR Dubna where part of this work was done.
S.T. acknowledges support by the
DFG cluster of excellence ``Origin and Structure of the Universe'',
by the Helmholtz International Center for FAIR (HIC for FAIR)
within the framework of the LOEWE program launched by
the state of Hesse via the Technical University Darmstadt and
by the Nuclear Astrophysics Virtual Institute of the Helmholtz
Association.


\begin{thebibliography}{90}
 
%\cite{Demorest:2010bx}
\bibitem{Demorest:2010bx}
  P.~Demorest et al., 
% T.~Pennucci, S.~Ransom, M.~Roberts, J.~Hessels,
  %``Shapiro delay measurement of a two solar mass neutron star,''
  Nature {\bf 467}, 1081 (2010).
%  [arXiv:1010.5788 [astro-ph.HE]].

%\cite{Ozel:2010bz}
\bibitem{Ozel:2010bz}
  F.~\"Ozel et al., 
% D.~Psaltis, S.~Ransom, P.~Demorest, M.~Alford,
  %``The Massive Pulsar PSR J1614-2230: Linking Quantum Chromodynamics, Gamma-ray Bursts, and Gravitational Wave Astronomy,''
  Astrophys. J. {\bf 724}, L199 (2010).
%  [arXiv:1010.5790 [astro-ph.HE]].

%\cite{Lattimer:2010uk}
\bibitem{Lattimer:2010uk}
  J.~M.~Lattimer, M.~Prakash,
  %``What a Two Solar Mass Neutron Star Really Means,''  
%  [arXiv:1012.3208 [astro-ph.SR]].
  arXiv:1012.3208.

%\cite{Weissenborn:2011qu}
\bibitem{Weissenborn:2011qu}
  S.~Weissenborn et al., 
% I.~Sagert, G.~Pagliara, M.~Hempel, J.~Schaffner-Bielich,
  %``Quark Matter In Massive Neutron Stars,''
    Astrophys. J. {\bf 740}, L14 (2011).
%  [arXiv:1102.2869 [astro-ph.HE]].

%\cite{Bonanno:2011ch}
\bibitem{Bonanno:2011ch}
  L.~Bonanno and A.~Sedrakian,
  %``Composition and stability of hybrid stars with hyperons and quark 
%color-superconductivity,''
%Astron. Astrophys. {\bf }, (2011).
% arXiv:1108.0559 [astro-ph.SR].
 arXiv:1108.0559.
  %%CITATION = ARXIV:1108.0559;%%

%\cite{Ozel:2006bv}
\bibitem{Ozel:2006bv} 
  F.~\"Ozel,
%``Soft equations of state for neutron-star matter ruled out by EXO 0748-676,''
  Nature {\bf 441}, 1115 (2006).
  %%CITATION = NATUA,441,1115;%%

%\cite{Alford:2006vz}
\bibitem{Alford:2006vz} 
  M.~Alford et al., 
% D.~Blaschke, A.~Drago, T.~Kl\"ahn, G.~Pagliara and  J.~Schaffner-Bielich,
  %``Quark matter in compact stars?,''
  Nature {\bf 445}, E7 (2007).
%  [astro-ph/0606524].
  %%CITATION = ASTRO-PH/0606524;%%

%\cite{Klahn:2006iw}
\bibitem{Klahn:2006iw} 
  T.~Kl\"ahn et al., 
% D.~Blaschke, F.~Sandin, C.~Fuchs, A.~Faessler, H.~Grigorian, 
%G.~R\"opke and J.~Tr\"umper,
  %``Modern compact star observations and the quark matter equation of state,''
  Phys.\ Lett.\ B {\bf 654}, 170 (2007).
%  [nucl-th/0609067].
  %%CITATION = NUCL-TH/0609067;%%

%\cite{Klahn:2011fb}
\bibitem{Klahn:2011fb}
  T.~Kl\"ahn, D.~Blaschke and R.~{\L}astowiecki,
  %``Compact Stars, Heavy Ion Collisions, and Possible Lessons For QCD at 
%Finite Densities,''
%  arXiv:1111.6889 [nucl-th].
  arXiv:1111.6889.
  %%CITATION = ARXIV:1111.6889;%%

%\cite{Pagliara:2007ph}
\bibitem{Pagliara:2007ph} 
  G.~Pagliara and J.~Schaffner-Bielich,
  %``Stability of CFL cores in Hybrid Stars,''
  Phys.\ Rev.\ D {\bf 77}, 063004 (2008).
%  [arXiv:0711.1119 [astro-ph]].
  %%CITATION = ARXIV:0711.1119;%%

%\cite{Ippolito:2007hn}
\bibitem{Ippolito:2007hn} 
  N.~Ippolito et al., 
% M.~Ruggieri, D.~Rischke, A.~Sedrakian and F.~Weber,
  %``Equilibrium sequences of non-rotating and rapidly rotating crystalline color superconducting hybrid stars,''
  Phys.\ Rev.\ D {\bf 77}, 023004 (2008).
%  [arXiv:0710.3874 [astro-ph]].
  %%CITATION = ARXIV:0710.3874;%%


%\cite{Blaschke:2010vj}
\bibitem{Blaschke:2010vj} 
  D.~Blaschke, J.~Berdermann and R.~{\L}astowiecki,
  %``Hybrid neutron stars based on a modified PNJL model,''
  Prog.\ Theor.\ Phys.\ Suppl.\  {\bf 186}, 81 (2010).
%  [arXiv:1009.1181 [nucl-th]].
  %%CITATION = ARXIV:1009.1181;%%


%\cite{Baldo:1999rq}
\bibitem{Baldo:1999rq} 
  M.~Baldo, G.~F.~Burgio and H.~J.~Schulze,
  %``Hyperon stars in the Brueckner-Bethe-Goldstone theory,''
  Phys.\ Rev.\ C {\bf 61}, 055801 (2000).
%  [nucl-th/9912066].
  %%CITATION = NUCL-TH/9912066;%%

%\cite{Baldo:2003vx}
\bibitem{Baldo:2003vx} 
  M.~Baldo, G.~F.~Burgio and H.~-J.~Schulze,
  %``Neutron star structure with hyperons and quarks,''
  astro-ph/0312446.
  %%CITATION = ASTRO-PH/0312446;%%

%\cite{Hofmann:2000mc}
\bibitem{Hofmann:2000mc}
  F.~Hofmann, C.~M.~Keil and H.~Lenske,
  %``Application of the density dependent hadron field theory to neutron  star
  %matter,''
  Phys.\ Rev.\  C {\bf 64}, 025804 (2001).
%  [arXiv:nucl-th/0008038].
  %%CITATION = PHRVA,C64,025804;%%

%\cite{Weissenborn:2011ut}
\bibitem{Weissenborn:2011ut}
  S.~Weissenborn, D.~Chatterjee and J.~Schaffner-Bielich,
  %``Hyperons and massive neutron stars: vector repulsion and SU(3) symmetry,''
%  arXiv:1112.0234 [astro-ph.HE].
   arXiv:1112.0234.
 %%CITATION = ARXIV:1112.023


%\cite{Typel:2009sy}
\bibitem{Typel:2009sy}
  S.~Typel et al.,
% G.~R\"{o}pke, T.~Kl\"{a}hn, D.~Blaschke and H.~H.~Wolter,
  %``Composition and thermodynamics of nuclear matter with light clusters,''
  Phys.\ Rev.\  C {\bf 81}, 015803 (2010).
%  [arXiv:0908.2344 [nucl-th]].
  %%CITATION = PHRVA,C81,015803;%%

%\cite{Blaschke:2005uj}
\bibitem{Blaschke:2005uj}
  D.~Blaschke et al., 
% S.~Fredriksson, H.~Grigorian, A.~M.~\"Oztas, F.~Sandin,
  %``The Phase diagram of three-flavor quark matter under compact star 
%constraints,''
  Phys.\ Rev.\  {\bf D72 } (2005)  065020.
%  [hep-ph/0503194].


%\cite{Grigorian:2006qe}
\bibitem{Grigorian:2006qe}
  H.~Grigorian,
  %``Parametrization of a nonlocal, chiral quark model in the instantaneous 
%three-flavor case: Basic formulas and tables,''
  Phys.\ Part.\ Nucl.\ Lett.\  {\bf 4}, 223 (2007).
%  [hep-ph/0602238].

%\cite{Alford:2004pf}
\bibitem{Alford:2004pf} 
  M.~Alford et al., 
%M.~Braby, M.~W.~Paris and S.~Reddy,
  %``Hybrid stars that masquerade as neutron stars,''
  Astrophys.\ J.\  {\bf 629}, 969 (2005).
%  [nucl-th/0411016].
  %%CITATION = NUCL-TH/0411016;%%

%\cite{Blaschke:2008br}
\bibitem{Blaschke:2008br} 
  D.~Blaschke et al., 
%F.~Sandin, T.~Kl\"ahn and J.~Berdermann,
  %``Sequential deconfinement of quark flavors in neutron stars,''
  Phys.\ Rev.\ C {\bf 80}, 065807 (2009).
%  [arXiv:0807.0414 [nucl-th]].
  %%CITATION = ARXIV:0807.0414;%%

\end{thebibliography}
\end{document}